\documentclass[12pt,twoside]{article}
\usepackage{epsfig}
\usepackage{graphicx}
\usepackage{amssymb}
\usepackage{url}
\usepackage{amsfonts,amsbsy}
\usepackage{amsmath}
\usepackage{authblk} 

\title{Decomposition of Electromagnetic Q and P Media}
\author[1]{I.V.\ Lindell\thanks{ismo.lindell@aalto.fi}}
\author[2]{A.\ Favaro} 
\affil[1]{Department of Radio Science and Engineering,}
\affil[ ]{Aalto University, School of Electrical Engineering, Espoo, Finland}
\affil[2]{The Blackett Laboratory, Department of Physics,}
\affil[ ]{Imperial College London, UK}
\date{\vspace{-20pt}}
\parskip=\medskipamount
\def\e{\begin{equation}} 
\def\f{\end{equation}} 
\def\ea{\begin{eqnarray}} 
\def\fa{\end{eqnarray}} 

\def\##1{{\mbox{\textbf{#1}}}}
\def\%#1{{\mbox{\boldmath $#1$}}}
\def\=#1{{\overline{\overline{\mathsf #1}}}}

\def\SE{{\mathbb E}}
\def\SF{{\mathbb F}}

\def\*{^{\displaystyle*}}

\def\.{\cdot}
\def\x{\times}

\def\ra{\rightarrow}

\def\Ra{\Rightarrow}

\def\l#1{\label{eq:#1}}
\def\r#1{(\ref{eq:#1})}
\def\am{\left(\begin{array}{c}}
\def\amm{\left(\begin{array}{cc}}
\def\ammm{\left(\begin{array}{ccc}}
\def\ammmm{\left(\begin{array}{cccc}}
\def\a{\end{array}\right)}

\def\A{\alpha}
\def\B{\beta}

\def\De{\Delta}
\def\E{\epsilon}
\def\g{\gamma}

\def\M{\mu}
\def\o{\omega}

\def\z{\zeta}

\def\Th{\Theta}

\def\VR{\varrho}

\def\ve{\%\varepsilon}

\def\tr{{\rm tr }}

\def\W{\wedge}
\def\WW{\displaystyle{{}^\wedge}\llap{${}_\wedge$}}

\def\J{\rfloor}
\def\L{\lfloor}
\def\JJ{\rfloor\rfloor}
\def\LL{\lfloor\lfloor}

\begin{document}

\maketitle

\begin{abstract}
Two previously studied classes of electromagnetic media, labeled as those of Q media and P media, are decomposed according to the natural decomposition introduced by Hehl and Obukhov. Six special cases based on either non-existence or sole existence of the three Hehl-Obukhov components, are defined for both medium classes.\end{abstract}

\section{Introduction}

Recent research on metamaterials and metaboundaries (see, e.g., \cite{Capolino}) has shown increased interest on electro\-magnetic media with properties not attained by simple isotropic media. The concept of bi-anisotropic medium was introduced by Cheng and Kong in 1968 \cite{Cheng68,Kong72} to describe the most general linear medium. In the representation involving Gibbsian vector fields, the medium equations of a bi-anisotropic medium can be written as
\e \am \#D\\ \#B\a = \amm \=\E & \=\xi\\ \=\z & \=\M\a\.\am \#E\\ \#H\a. \l{mediumeq}\f
The four medium dyadics involve $4\x9=36$ medium parameters in the most general case \cite{Kong,Methods}. 

The set of electromagnetic medium parameters can be most naturally decomposed in three invariant parts which aids in defining different classes of media. The decomposition requires four-dimensional formalism for the Maxwell equations in terms of differential forms. Following \cite{Deschamps,Difform}, we denote by $\%\Phi$ and $\%\Psi$ the 4D field two-forms which can be expanded in terms of the 3D (spatial) field two-forms $\#B,\#D$ and one-forms $\#E,\#H$ as
\e \%\Phi = \#B + \#E\W\ve_4,\ \ \ \ \%\Psi= \#D - \#H\W\ve_4, \f
in terms of which the Maxwell equations can be expressed in compact form as
\e \#d\W\%\Phi = 0,\ \ \ \#d\W\%\Psi=\%\g. \f
Here, $\%\g$ denotes the source three-form whose expansion in terms of 3D charge three-form $\%\VR$ and current two-form $\#J$ reads
\e \%\g = \%\VR -\#J\W\ve_4. \f
The one-form $\ve_4=c\#d t$ denotes the temporal component of the one-form basis $\ve_1,\ve_2,\ve_3,\ve_4$ while $\#d$ is the differentiation one-form. More details of the notation applied here can be found in \cite{Difform} or \cite{MDEM}. The medium equation \r{mediumeq} is represented by the simple linear relation between the field two-forms as
\e \%\Psi=\=M|\%\Phi, \l{mediumeq4}\f
where $\=M$ is a bidyadic, mapping two-forms to two-forms. It contains the information of the four 3D dyadics of \r{mediumeq} and can be represented as a $6\x6$ matrix in any chosen basis system. 

A most natural decomposition of $\=M$ in three parts was defined by Hehl and Obukhov \cite{Hehl} as 
\e \=M = \=M_1+ \=M_2+ \=M_3, \f
respectively labeled as principal, skewon and axion components of $\=M$. When making an affine transformation to the spacetime (correponding to stretching and rotating the space and setting it in uniform motion), the three parts are transformed individually. To define the Hehl-Obukhov decomposition, we can apply the double-contraction mapping \cite{MDEM}
\e \=I{}^{(4)T}\LL(\=M_1+\=M_2+\=M_3)^T = \=M_1- \=M_2+ \=M_3, \f
which defines the skewon component as
\e \=M_2 = \frac{1}{2}(\=M -\=I{}^{(4)T}\LL\=M{}^T). \l{M22}\f
The axion component is defined as containing the trace of the medium bidyadic,
\e \=M_3=\frac{1}{6}(\tr\=M) \=I{}^{(2)T},\f
where $\=I{}^{(2)}$ denotes the unit bidyadic for bivectors and its transpose to two-forms. Finally, the principal part of $\=M$ can be obtained as $\=M_1=\=M-\=M_2-\=M_3$ and we can write
\e \=M_1 = \frac{1}{2}(\=M+ \=I{}^{(4)T}\LL\=M{}^T) - \frac{1}{6}\tr\=M\ \=I{}^{(2)T}. \f 
The principal and skewon components are trace free, $\tr\=M_1=\tr\=M_2=0$. Applying the identity valid for any bidyadic $\=M$  \cite{MDEM},
\e \=I{}^{(4)T}\LL\=M{}^T = (\tr\=M)\=I{}^{(2)T} -(\=M\LL\=I)\WW\=I{}^T + \=M \f
to \r{M22}, the skewon part of $\=M$ can be expressed in terms of a trace-free dyadic $\=B_o\in\SE_1\SF_1$ in the form
\e \=M_2 =  (\=B_o\WW\=I)^T,\ \ \ \ \tr\=B_o=0, \l{05M2Bo}\f
with
\e \=B_o = \frac{1}{2}(\=M\LL\=I)^T- \frac{1}{4}(\tr\=M)\=I. \l{05BoM2}\f 
It can be shown that that if the medium bidyadic satisfies the double-contraction condition
\e \=M\LL\=I =0, \l{05M1LLI}\f
it equals its principal part, $\=M=\=M_1$ \cite{MDEM}. Thus, \r{05M1LLI} represents the condition of a medium without skewon and axion components. Definitions and basic properties of the various products $|$ $\W$, $\L$, $\J$ and the double products $||$, $\WW$, $\LL$ and $\JJ$ can be found in \cite{Difform} or \cite{MDEM}.

Defining the modified medium bidyadic by the contraction operation as
\e \=M_m=\#e_N\L\=M,\ \ \ \ \#e_N=\#e_{1234}=\#e_1\W\#e_2\W\#e_3\W\#e_4, \f
its skewon component can be shown to consist of the antisymmetric part of $\=M_m$,
\e \=M_{m2} = \frac{1}{2}(\=M_m - \=M{}_m^T), \f
while the sum of the principal and axion components corresponds to the symmetric part of $\=M_m$. 

The purpose of this study is to consider the Hehl-Obukhov decomposition of two classes of media whose medium bidyadic $\=M$ can be expressed in terms of a dyadic $\=P$ mapping vectors to vectors or a dyadic $\=Q$ mapping one-forms to vectors. Each of them involves $4\x4=16$ parameters in general. Such media have been introduced in the past and labeled as Q media \cite{Difform,416,421} and P media \cite{P} for brevity. 

The class of Q media can be characterized as containing ``ordinary media", e.g., media representing polarizable dielectric and magnetic materials \cite{ThesisAlberto}. Recently, Q media have been considered as extensions to the usual electromagnetic response of vacuum \cite{Ni2015}. In this context, Q media allow for the vacuum response to include a finite skewon part without disrupting the light cone. Thus, the propagation of electromagnetic waves remains free of birefringence, as required by the observational evidence. For instance, polarisation observations of gamma-ray bursts specify the vanishing of cosmic birefringence with $10^{-38}$ accuracy \cite{Ni2015}. Gravitational theories in which the metric is replaced by an asymmetric tensor (that one could identify with $\=Q\hspace{1pt}$) have been studied in the past \cite{Einstein,Schroedinger}, but are nowadays deemed not viable. 

It is known that P media have quite special properties, e.g., as an example of media with no dispersion equation \cite{NDE}. Their realization in terms of metamaterials may, however, still take some effort. It has turned out that novel sets of boundary conditions can be obtained at the interface of certain P media \cite{PEMC,IB,DB,SHDB}. Since such boundary conditions have recently gained important applications, see, e.g., \cite{Zhang,Yaghjian08,Yaghjian10}, their practical realization has also created interest \cite{Caloz,Elmaghrabi,Zaluski11,Zaluski14}. 

It is important to note that, in this work, the medium parameters are assumed to be complex numbers. Hence, the medium bidyadic $\=M$ is complex, as opposed to real. Here, we allow the engineering practice of assuming time dependence $\exp(j\o t)$ for the fields to avoid unnecessary complications in the analysis. Thus, all quadratic equations, both for scalars and for dyadics, are solved over the complex numbers. Nevertheless, operating strictly within the real numbers is possible. This approach is used, for example, in \cite{Hehl, ThesisAlberto}. 

\section{Decomposition of Q media}

The class of Q media is defined by modified medium bidyadics of the form \cite{Difform}
\e \=M_m = \=Q{}^{(2)}, \f
where $\=Q$ is a dyadic mapping one-forms to vectors. Let us expand 
\e \=Q = \=S + \#A\L\=I{}^T, \f
where $\=S$ is the symmetric part of $\=Q$ and the antisymmetric part can be expressed in terms of a bidyadic $\#A$. The number of parameters is $10 \ (\=S)\ +6\ (\#A)\ =16\ (\=Q)$. Writing
\e \=M_m = \=S{}^{(2)} + \=S\WW(\#A\L\=I{}^T) + (\#A\L\=I{}^T)^{(2)}, \l{QMm}\f
the last term can be expanded applying the identity \cite{Difform}
\e (\#A\L\=I{}^T)^{(2)} = \#A\#A - \frac{1}{2}\ve_N|(\#A\W\#A)\#e_N\L\=I{}^{(2)T}. \f
The middle term in \r{QMm} is the antisymmetric part of $\=M_m$ and, hence, represents the skewon component, while the other two terms are symmetric. The trace of $\ve_N\L\=S{}^{(2)}$ for any symmetric dyadic $\=S$ vanishes, which is seen by expanding $\=S=\sum_{i=1}^{4} \#s_i\#s_i$, whence
\e \tr(\ve_N\L\=S{}^{(2)}) = \frac{1}{2}\sum_i\sum_j (\ve_N\L(\#s_i\W\#s_j))|(\#s_i\W\#s_j) = \frac{1}{2}\sum_i\sum_j\ve_N|(\#s_i\W\#s_j\W\#s_i\W\#s_j)=0. \f
Thus, the trace of the Q-medium bidyadic becomes
\e \tr\=M = \tr(\ve_N\L\#A\#A) - \frac{1}{2}\ve_N|(\#A\W\#A)\tr\,\=I{}^{(2)T} = -2\ve_N|(\#A\W\#A). \f
It vanishes when the bivector $\#A$ is simple, satisfying $\#A\W\#A=0$.

Applying the identity \r{BoI}, the skewon component of the Q-medium \r{QMm} can be written in the form \r{05M2Bo} as
\e \=M_{m2} = \=S\WW(\#A\L\=I{}^T) = \#e_N\L(\=B{}^T_o\WW\=I{}^T),\f
as defined by the dyadic
\e \=B{}^T_o=\ve_N\L(\#A\W\=S) = (\ve_N\L\#A)\L\=S. \l{QBoT}\f
In conclusion, the Hehl-Obukhov components of the Q medium bidyadic $\=M$ are
\ea \=M_3 &=& -\frac{1}{3}\ve_N|(\#A\W\#A)\=I{}^{(2)T}, \l{M3}\\
 \=M_2 &=& (\ve_N\L(\#A\W\=S))\WW\=I{}^T, \l{M2}\\
 \=M_1 &=& \ve_N\L(\=S{}^{(2)} + \#A\#A) - \frac{1}{6}\ve_N|(\#A\W\#A)\=I{}^{(2)T}. \l{M1}\fa
For comparable results, see \cite{ThesisAlberto,Ni2015}.

\section{Special Cases of Q Media}

It appears that there are six basic special cases of the Q medium, based on one or two missing components of the Hehl-Obukhov decomposition. Let us examine these cases separately. 

\subsection{No axion component}

The case of no axion component in the Q medium, $\=M_3=0$, requires from \r{M3} that the bivector $\#A$ satisfy the condition $\#A\W\#A=0$. Thus, the antisymmetric part of the $\=Q$ dyadic is defined by a simple bivector of the form $\#A=A\#a\W\#b$. The modified medium bidyadic has then the form
\ea \=M_m &=& (\=S + A(\#a\W\#b)\L\=I{}^T)^{(2)}\nonumber\\
&=&\=S{}^{(2)} - A\#e_N\L((\ve_N\L(\#a\W\#b\W\=S))\WW\=I{}^T) + A^2(\#a\W\#b)(\#a\W\#b).\fa
In the special case $\#A=0$, i.e., when a Q medium is defined by a symmetric dyadic $\=Q=\=S$, the medium bidyadic $\=M_m=\=S{}^{(2)}$ has only a principal component.

\subsection{No skewon component}

From \r{05M2Bo} and \r{QBoT} it follows that the skewon component of a Q medium bidyadic vanishes for
\e \=B{}^T_o = \%\Th\L\=S=0,\ \ \ \%\Th=\ve_N\L\#A. \l{M20}\f
Multiplying \r{M20} by $(\#A\L\=I{}^T)|$ and using the identity \r{PhiLAA}, considered in the Appendix, we obtain 
\e \#A\L\=B{}_o^T = \#A\L(\%\Th\L\=S) =  -\frac{1}{2}\ve_N|(\#A\W\#A)\=S=0, \f
whence either $\=S=0$ or the bivector $\#A$ is simple, satisfying $\#A\W\#A=0$. In the previous case we have
\e \=M_m= (\#A\L\=I{}^T)^{(2)} = \#A\#A -\frac{1}{2}(\#A\W\#A)\L\=I{}^{(2)T}. \l{MmAA}\f

Assuming $\=S\not=0$ we must have $\#A\W\#A=0$, which case corresponds to a pure-principal Q medium. For $\#A=0$ we have $\=M=\=S{}^{(2)}$. Assuming $\#A=A\#a\W\#b\not=0$, the condition \r{M20} can be written as
\e \ve_N\L(\#a\W\#b\W\=S)=0\ \ \ \Ra\ \ \ \ \#a\W\#b\W\=S=0, \l{abS}\f
which requires that the symmetric dyadic must be of the form
\e \=S= S_{aa}\#a\#a+ S_{ab}(\#a\#b+\#b\#a) + S_{bb}\#b\#b, \l{S}\f
satisfying
\e \=S{}^{(2)} = (S_{aa}S_{bb}-S_{ab}^2)(\#a\W\#b)(\#a\W\#b),  \l{S2}\f
and
\e \=Q = S_{aa}\#a\#a+ (S_{ab}-A)\#a\#b+(S_{ab}+A)\#b\#a + S_{bb}\#b\#b. \f 
In this case the modified medium bidyadic must have the quite restricted form of rank 1,
\e \=M_m = (\=S+ \#A\L\=I{}^T)^{(2)} = \A(\#a\W\#b)(\#a\W\#b),\ \ \ \ \A=S_{aa}S_{bb}-S_{ab}^2+A^2. \l{Mm20}\f
To conclude, for a skewon free Q medium we have two possibilities, either the pure-principal medium $\=M_m=\=S{}^{(2)}$ or the principal-axion medium of \r{MmAA}. The pure-principal medium defined by \r{Mm20} coincides with \r{MmAA} for $\#A\W\#A=0$. These findings agree with Table 3.2 of \cite{ThesisAlberto}, and generalise it somewhat. As a matter of fact, $\=Q$ is not demanded to be invertible, here.

\subsection{No principal component}

Setting $\=M_{m1}=0$ in  \r{M1} we obtain the following condition for the modified Q-medium bidyadic with no principal component:
\e \=S{}^{(2)} =- \#A\#A + \frac{1}{6}\ve_N|(\#A\W\#A)\#e_N\L\=I{}^{(2)T}. \l{Mm1}\f
It can be shown that $\=S{}^{(2)}=0$ leads to $\#A=0$, whence $\=M=0$. For $\=S{}^{(2)}\not=0$ we can operate each side of \r{Mm1} as
\e \=S{}^{(2)}|(\ve_N\L\=S{}^{(2)}) = \#e_N\L(\ve_N\ve_N\LL\=S{}^{(2)})|\=S{}^{(2)} = \De_S\#e_N\L\=I{}^{(2)T} \f
and
$$ (-\#A\#A + \frac{1}{6}\ve_N|(\#A\W\#A)\#e_N\L\=I{}^{(2)T})|(\ve_N\L(-\#A\#A + \frac{1}{6}\ve_N|(\#A\W\#A)\#e_N\L\=I{}^{(2)T})) =$$
\e= (\ve_N|(\#A\W\#A))(\frac{2}{3}\#A\#A + \frac{1}{36}\ve_N|(\#A\W\#A)\#e_N\L\=I{}^{(2)T}), \f
with
\e \De_S=\ve_N\ve_N||\=S{}^{(4)}, \f
we obtain
\e (\De_S-\frac{1}{36}(\ve_N|(\#A\W\#A))^2)\#e_N\L\=I{}^{(2)T} = \frac{2}{3}(\ve_N|(\#A\W\#A))\#A\#A. \f
Since $\#e_N\L\=I{}^{(2)T}$ is of full rank and $\#A\#A$ is not, both sides of this equation must vanish, whence $\#A\W\#A=0$ and  $\De_S=0$. Thus, the axion component of the medium bidyadic must vanish, $\#A=A\#a\W\#b$, and the dyadic $\=S$ cannot be of full rank. From \r{Mm1} we now have
\e \=S{}^{(2)} =- \#A\#A = A^2\#a\#b\WW\#b\#a, \f
whence for $\#A\not=0$ the dyadic $\=S$ must be of the form
\e \=S = S_{aa}\#a\#a+ S_{ab}(\#a\#b+\#b\#a) + S_{bb}\#b\#b,\ \ \ S_{aa}S_{bb}-S_{ab}^2=-A^2.\f
Because of
\e \=M_m =\=S\WW(\#A\L\=I{}^{(2)}) = A\=S\WW(\#b\#a-\#a\#b)=0, \f
a Q medium without a principal component does not exist.

\subsection{Pure-principal media}

Because a Q medium does not exist without a principal part, there are no pure-axion or pure-skewon Q media. To define possible pure-principal Q media, assuming $\#A=0$ both axion and skewon components vanish, whence $\=M_m=\=S{}^{(2)}$ is pure principal for any symmetric dyadic $\=Q=\=S$. 
For $\#A=A\#a\W\#b\not=0$, vanishing skewon component requires the condition \r{abS}. In this case the dyadic $\=S$ must be of the form \r{S} and the resulting modified medium bidyadic of the simple form \r{Mm20}. The same form is valid also in the case $\=S=0$ with $\#A\W\#A=0$.

\subsection{Summary: Special cases of Q media}

From the above results we conclude that the principal component is essential for a Q medium, because without a principal component the medium does not exist. Thus, the only special cases are principal-skewon, principal-axion and pure-principal Q media.

\begin{itemize}

\item $\=M_1\not=0,\ \=M_2=0,\ \=M_3=0$, pure-principal Q media.
\e \=Q = \=S,\ \ \ \ \=M_m = \=S{}^{(2)},\f
or
\e \=Q=S_{aa}\#a\#a+ S_{ab}(\#a\#b+\#b\#a) + S_{bb}\#b\#b + A(\#a\W\#b)\L\=I{}^T,\ \ \ \ \=M_m = M(\#a\W\#b)(\#a\W\#b) \f 
with $M=S_{aa}S_{bb}-S_{ab}^2 + A^2$.

\item $\=M_1\not=0,\ \=M_2=0,\ \=M_3\not=0$, skewon-free Q media.
\e \=Q= \#A\L\=I{}^T,\ \ \ \ \ \=M_m = (\#A\L\=I)^{(2)} = \#A\#A - \frac{1}{2}(\#A\W\#A)\L\=I{}^T. \f

\item $\=M_1\not=0,\ \=M_2\not=0,\ \=M_3=0$, axion-free Q media.
\e \=Q = \=S + A(\#a\W\#b)\L\=I{}^T,\ \ \ \ \=M_m = \=S{}^{(2)} -A\#e_N\L((\ve_N\L(\#a\W\#b\W\=S))\WW\=I{}^T)+ A^2(\#a\W\#b)(\#a\W\#b).  \f

\item $\=M_1=0\ \ \Ra\ \ \=M=0$ (skewon-axion, pure-skewon and pure-axion Q media are not possible).

\end{itemize}

\section{Decomposition of P media}

The class of P media is defined by medium bidyadics of the form
\e \=M = \=P{}^{(2)T} \l{MP2T},\f
involving a dyadic $\=P$ mapping vectors to vectors \cite{P}. P media are somewhat strange since there is no dispersion equation to limit the choice of the $\#k$ vector of a plane wave \cite{MDEM}. Next, the Hehl-Obukhov decomposition of these media is performed. The ensuing formulae for the principal, skewon and axion parts are analogous to those in \cite{ThesisAlberto}. 

Extracting the trace-free part $\=P_o$ from the dyadic $\=P$ as
\e \=P = \=P_o + P\=I,\ \ \ P=\frac{1}{4}\tr\=P,\ \ \ \tr\=P_o=0, \f
the P-medium bidyadic can be expanded as
\e \=M =P^2\=I{}^{(2)T} + P(\=P_o\WW\=I)^T +\=P{}_o^{(2)T}. \l{MP}\f
Since the first term has only an axion component and the second term has only a skewon component, it is clear that the $P$ parameter does not have any effect on the principal component of the P medium. The trace of the last term yields
\e \tr\=P{}_o^{(2)T} = \frac{1}{2}((\tr\=P_o)^2-\tr\=P{}_o^2)=-\frac{1}{2}\tr\=P{}_o^2, \f
whence the axion component of the P-medium bidyadic has the form
\e \=M_3 = \frac{1}{6}\tr\=M\ \=I{}^{(2)T} =(P^2-\frac{1}{12}\tr\=P{}_o^2)\=I{}^{(2)T}. \l{PM3}\f
The skewon component of any medium is specified by \r{05M2Bo} and \r{05BoM2}. Accordingly, the skewon component of a P medium is given by $\=M_2 =  (\=B_o\WW\=I)^T$ with
\ea \=B_o &=& \frac{1}{2}\=P{}^{(2)}\LL\=I{}^T- \frac{1}{4}(\tr\=P{}^{(2)})\=I \nonumber\\
 &=& -\frac{1}{2}\=P{}^2_o +P\=P_o + \frac{1}{8}(\tr\=P_o{}^2)\=I{}\nonumber\\
&=& - \frac{1}{2}(\=P_o -P\=I)^2 + \frac{1}{2}(P^2 + \frac{1}{4}\tr\=P{}_o^2)\=I\nonumber\\
&=& - \frac{1}{2}((\=P_o -P\=I)^2 - \frac{1}{4}\tr(\=P_o-P\=I)^2\=I). \l{PBo}\fa
More explicitly, the $\=B_o$ dyadic is equal to the trace-free part of the dyadic $-(\=P_o-P\=I)^2/2$.  

Finally, the principal component of the P-medium bidyadic can be expressed as
\ea \=M_1 &=& \=M - \=M_2 - \=M_3 \nonumber\\
&=& \=P{}_o^{(2)T}  + \frac{1}{2}\=P{}_o^{2T}\WW\=I{}^T -\frac{1}{6}(\tr\=P{}_o^2)\=I{}^{(2)T}\nonumber\\
&=& \frac{1}{2}(\=P_o\WW\=I)^{2T} -\frac{1}{6}(\tr\=P{}_o^2)\=I{}^{(2)T}\nonumber\\
&=&\frac{1}{2}((\=P_o\WW\=I)^2 - \frac{1}{6}\tr(\=P_o\WW\=I)^2\ \=I{}^{(2)})^T, \l{PM1}\fa
where we have applied the expansions
\ea (\=P_o\WW\=I)^2 &=& \=P_o\WW\=P_o + \=P{}_o^2\WW\=I, \nonumber\\
\tr(\=P_o\WW\=I)^2 &=&  (\tr\=P_o)^2-\tr\=P{}_o^2 + 3\tr\=P{}_o^2=2 \tr\=P{}_o^2, \fa
see \cite{MDEM}, p.348. The expression \r{PM1} can be interpreted so that the principal part of the P-medium bidyadic equals the trace-free part of the bidyadic $(\=P_o\WW\=I)^{2T}/2$. Again, we note that the principal part of the P-medium bidyadic $\=M_1$ is independent of the parameter $P$. To check the expression \r{PM1}, we can expand
\ea \=M_1\LL\=I &=& \=P{}_o^{(2)T}\LL\=I  + \frac{1}{2}(\=P{}_o^{2T}\WW\=I{}^T)\LL\=I -\frac{1}{2} (\tr\=P{}_o^2)\=I{}^T \nonumber\\
&=& -\=P{}_o^{2T} + \frac{1}{2}((\tr\=P{}_o^2)\=I{}^T+ 2\=P{}_o^{2T}) -\frac{1}{2} (\tr\=P{}_o^2)\=I{}^T , \fa
which vanishes identically. Thus, \r{PM1} indeed represents a pure-principal bidyadic.

\section{Special Cases of P Media}

Let us consider the six special cases of the P medium.

\subsection{No axion component}

From \r{PM3} the condition for axion-free P medium becomes
\e P = \frac{1}{2}\sqrt{\frac{1}{3}\tr\=P{}_o^2}, \l{noaxion}\f
with either branch of the square root. The medium bidyadic has then the form
\ea \=M &=& (\=P_o + \frac{1}{2}\sqrt{\frac{1}{3}\tr\=P{}_o^2}\ \=I)^{(2)T} \nonumber\\
&=& \=P{}_o^{(2)T} + \frac{1}{2}\sqrt{\frac{1}{3}\tr\=P{}_o^2}\ (\=P_o\WW\=I)^T+\frac{1}{12}(\tr\=P{}_o^{2T})\=I{}^{(2)T}. \fa
Such a bidyadic satisfies $\tr\=M=0$ for any trace-free dyadic $\=P_o$, as can be easily verified.

\subsection{No skewon component}

Any P medium with no skewon component must satisfy $\=B_o=0$ according to  \r{05BoM2}. Applying \r{PBo}, we obtain the condition 
\e (\=P_o -P\=I)^2 = \A{}^2\ \=I, \l{PoPI2}\f
with
\e \A^2 = \frac{1}{4}\tr(\=P_o -P\=I)^2 = \frac{1}{4}\tr\=P{}^2_o +P^2. \f 
Let us consider the two cases $\A=0$ and $\A\not=0$ separately.

\subsubsection{$\A=0$}

Assuming
\e 4\A{}^2 = \tr\=P{}_o^2 + 4P^2 =0, \f
from \r{PoPI2} the dyadic $\=P_o$ must have the form
\e \=P_o =P\=I +\=N, \f  
where $\=N$ is any nilpotent dyadic satisfying
\e \=N{}^2=0.\f 
One can show \cite{MDEM} that such a nilpotent dyadic can be at most of rank 2 and it can be expressed as
\e \=N= \#a_1\%\B_1+ \#a_2\%\B_2, \l{N2}\f
where the two vectors and one-forms satisfy the orthogonality conditions
\e \#a_1|\%\B_1=\#a_1|\%\B_2=\#a_2|\%\B_1=\#a_2|\%\B_2=0. \l{aiAj}\f
Because of $\tr\=N=0$, we have $P = \frac{1}{4}(\tr\=P_o -\tr\=N)=0$ and $\=P_o = \=N = \#a_1\%\B_1+ \#a_2\%\B_2$. Moreover,
\e \=P{}_o^2=0,\ \ \ \ \tr\=P{}_o^2=0. \f
As a result, the skewon-free condition \r{PoPI2} is obviously satisfied for $\A=0$. In this case the medium bidyadic must be of the form
\e \=M = \=P{}_o^{(2)T} = \%\B_1\#a_1\WW\%\B_2\#a_2, \l{PoN} \f
restricted by \r{aiAj}. To check this result, we expand
\e \=M\LL\=I=\tr(\%\B_1\#a_1)\%\B_2\#a_2 + \tr(\%\B_2\#a_2)\%\B_1\#a_1 - (\%\B_1\#a_1)|(\%\B_2\#a_2) - (\%\B_2\#a_2)|(\%\B_1\#a_1)=0. \f
Because $\=M$ satisfies \r{05M1LLI}, there is only a principal component and, thus, no skewon component.

\subsubsection{$\A\not=0$}

Considering the second case $\A\not=0$, or
\e  \tr\=P{}^2_o +4P^2\not=0, \l{Anot0}\f 
from \r{PoPI2} the dyadic $\=P_o-P\=I$ is proportional to a unipotent dyadic \cite{MDEM} (an involution \cite{Prasolov}). Expressing 
\e \=P_o- P\=I = \A(\=I-2\=\Pi) = \A(\=\Pi'-\=\Pi), \l{4P2}\f
and inserting in \r{PoPI2}, the condition is satisfied if $ \=\Pi$ is a projection dyadic, and $\=\Pi'$ the complementary projection dyadic, satisfying
\e \=\Pi+\=\Pi'=\=I,\ \ \ \ \ \=\Pi{}^2=\=\Pi,\ \ \ \ \=\Pi'{}^2=\=\Pi'. \f
One can show \cite{MDEM} that the trace of a projection dyadic equals its rank $p$. In consequence, $\tr\=\Pi=p$ is an integer between $0$ and $4$. Moreover, it is easy to check that $\tr\=\Pi'=p'=4-p$. Actually, $\=\Pi$ acts as the unit dyadic in a $p$-dimensional subspace of the vector space and $\=\Pi'$ acts as the unit dyadic in the complementary subspace of dimension $p'$. Taking the trace of \r{4P2} yields a relation between the parameters $P$ and $\A$,
\e 4P = \A(p-p') = \A(2p-4) = \A(4-2p').\l{pp'}\f
Because the sign of $\A$ is of no concern in \r{PoPI2}, it is sufficient to consider the three basic subcases, $p=0,1,2$. 

\begin{itemize}

\item Assuming $p=\tr\=\Pi=0$ and $p'=\tr\=\Pi'=4$ corresponds to $\=\Pi=0$ and $\=\Pi'=\=I$, whence from \r{pp'} we have that $P=-\A$ and from \r{4P2} that $\=P_o=(P+\A)\=I =0$. The condition \r{PoPI2} is now satisfied for $\=P=P\=I$ for any $P$. In this case the skewonless P medium equals the pure-axion medium.

\item Assuming $p=1$ and $p'=3$, \r{pp'} yields $2P=-\A$, whence from \r{4P2} we obtain 
\e \=P_o = P\=I -2P(\=I-2\=\Pi) = -P(\=I-4\=\Pi), \f
the right side of which is trace free. In this case we have
\e \=M = (\=P_o+ P\=I)^{(2)T} = (4P\=\Pi)^{(2)T} =0, \f
since $\=\Pi$ is of rank 1. Thus, there is no P medium correponding to $p=1$ . 

\item Finally, we assume that $p=p'=2$. Substituting this into \r{pp'} yields $P = 0$, whence
\e\=P= \A(\=\Pi'-\=\Pi) = \A(\=I - 2\=\Pi)=\=P_o.\f
One can verify that \r{PoPI2} is fulfilled for any scalar $\A$. To summarise, $\=P$ is proportional to an arbitrary trace-free unipotent dyadic.
We can define reciprocal bases of vectors $\{\#e_i\}$ and one-forms $\{\ve_j\}$ so that we can write
\ea \=\Pi &=& \#e_1\ve_1+ \#e_2\ve_2, \\ 
\=\Pi' &=& \#e_3\ve_3 + \#e_4\ve_4, \fa
whereby the $\=P$ dyadic becomes
\e \=P = \A(\#e_3\ve_3 + \#e_4\ve_4-\#e_1\ve_1- \#e_2\ve_2). \f
The medium bidyadic has now the form
\e \=M = -\A^2(\=I{}^{(2)T} -2\ve_{12}\#e_{12} - 2\ve_{34}\#e_{34}). \l{pureprinc}\f  
Because the modified medium bidyadic 
\e \=M_m = M(\#e_N\L\=I{}^{(2)T} -2\#e_{34}\#e_{12}- 2\#e_{12}\#e_{34}) \f
is symmetric, such a P medium does not have a skewon component. In this case the axion and principal components are both nonzero.
\end{itemize}
To conclude, there are three kinds of skewonless P-medium bidyadics, the pure-axion medium bidyadics, pure-principal bidyadics of the rank-1 form \r{PoN}, and principal-axion bidyadics of the form \r{pureprinc}.

\subsection{No principal component}

The condition for no principal component in a P medium is obtained from \r{PM1} as 
\e (\=P_o\WW\=I)^2 = \=P_o\WW\=P_o + \=P{}_o^2\WW\=I = \A^2 \=I{}^{(2)},\ \ \ \ \A = \sqrt{\frac{1}{3}\tr\=P{}_o^2}. \l{PoWWI}\f
There is no restriction concerning the scalar $P$, which affects only the axion and skewon components of the P medium. Operating \r{PoWWI} by $\LL\=P{}_o^T$ as
\ea 0 &=&  (\=P_o\WW\=P_o)\LL\=P{}_o^T + (\=P{}_o^2\WW\=I)\LL\=P{}_o^T -\frac{1}{3}\tr\=P{}_o^2\, \=I{}^{(2)}\LL\=P{}_o^T \nonumber\\
&=& 2((\tr\=P{}_o^2)\=P_o - \=P{}_o^3) + (\tr\=P{}_o^3)\=I - 2\=P{}_o^3 +\frac{1}{3}(\tr\=P{}_o^2)\=P_o, \fa
yields the condition
\e 4\=P{}_o^3-\frac{7}{3}(\tr\=P{}_o^2)\=P_o - (\tr\=P{}_o^3)\=I =0. \l{Po3}\f
Multiplying this by $\=P_o|$ we obtain
\e \=P{}_o^4=\frac{7}{12}(\tr\=P{}_o^2)\=P{}^2_o +\frac{1}{4} (\tr\=P{}_o^3)\=P_o , \l{Po4}\f
whose trace yields the relation
\e \tr\=P{}_o^4 = \frac{7}{12}(\tr\=P{}_o^2)^2. \l{trPo4}\f
Substituting \r{Po4} in the Cayley-Hamilton equation
\e \=P{}_o^4 -\tr\=P_o\ \=P{}_o^{3}+ \tr\=P{}_o^{(2)}\=P{}_o^2 -\tr\=P{}_o^{(3)}\=P_o +\tr\=P{}_o^{(4)}\=I=0, \l{CH1}\f
together with $\tr\=P_o=0$ and
\ea \tr\=P{}_o^{(2)} &=& -\frac{1}{2}\tr\=P{}_o^2,\\
 \tr\=P{}_o^{(3)} &=& \frac{1}{3}\tr\=P{}_o^3,\\
 \tr\=P{}_o^{(4)} &=& \frac{1}{8}((\tr\=P{}_o^2)^2 - 2\tr\=P{}_o^4)= -\frac{1}{48}(\tr\=P{}_o^2)^2, \fa
we obtain the condition 
\e (\tr\=P{}_o^2)\=P{}_o^2 -(\tr\=P{}_o^3)\=P_o - \frac{1}{4}(\tr\=P{}_o^2)^2\=I=0. \l{trP2P2}\f
Multiplying this by $\=P{}_o^2|$, taking the trace and applying \r{trPo4} leads to the relation
\e 3(\tr\=P{}_o^3)^2 = (\tr\=P{}_o^2)^3, \l{tr32}\f
whence the parameter $\A$ can be expressed as
\e \A = \sqrt{\frac{1}{3}\tr\=P{}_o^2} =  \sqrt{\frac{(\tr\=P{}_o^3)^2}{(\tr\=P{}_o^2)^2}}= \frac{\tr\=P{}_o^3}{\tr\=P{}_o^2}, \f 
Thus, \r{trP2P2} can be written in the form
\e \=P{}_o^2=\A\=P_o +\frac{3}{4}\A^2 \=I. \l{Po22}\f
At this point we can note that $\tr\=P{}_o^2\ra0$ corresponds to $\A\ra0$, whence $\=P_o\ra0$ and the P medium becomes a pure axion medium. 

Inserting \r{Po22} in \r{PoWWI}, the condition for the trace-free dyadic $\=P_o$ defining a medium with no principal component becomes
\ea \=P{}_{o}^{(2)} + \frac{1}{2}\=P{}_o^2\WW\=I - \frac{1}{2}\A^2\=I{}^{(2)} 
&=& \=P{}_{o}^{(2)} + \frac{1}{2}(\A\=P{}_o+\frac{3}{4}\A^2 \=I)\WW\=I - \frac{1}{2}\A^2\=I{}^{(2)} \nonumber\\
&=& \=P{}_{o}^{(2)} + \frac{\A}{2}\=P{}_o\WW\=I + \frac{\A^2}{4}\=I{}^{(2)} \nonumber\\
&=& (\=P_{o} + \frac{\A}{2}\=I)^{(2)} =0. \fa
Because the dyadic in brackets must be of rank one, we can write the general form for the dyadic $\=P_o$ as
\e \=P_o = A\=I +\#a\%\B ,\ \ \ \ \ A= -\frac{\#a|\%\B}{4}. \l{PoaB}\f
Adding the term $P\=I$ yields the dyadic $\=P$ which can be expressed in the form
\e \=P = P'\=I + \#a\%\B,\ \ \ \ P'=P+A. \f
To check that this leads to no principal component of the P medium, we can expand
\ea \=M &=& (P'\=I+ \#a\%\B)^{(2)T}  \nonumber\\
&=& P'{}^2\=I{}^{(2)T} + P'(\=I\WW\#a\%\B)^T\nonumber\\
&=& (\=B\WW\=I)^T, \l{Mprincc}\fa 
with
\e \=B = \frac{P'{}^2}{2}\=I +P'\#a\%\B. \l{Mprincc1}\f
Since the medium bidyadic \r{Mprincc} is of the skewon-axion form, it contains no principal part.

\subsection{Pure-axion media}

For a P medium consisting of an axion component only, the $\=P$ dyadic must satisfy
\e \=P{}^{(2)} = M\=I{}^{(2)},\ \ \ M\not=0.\l{P2MI}\f
Operating \r{P2MI} by $\LL\=I{}^T$ yields
\e (\tr\=P)\=P - \=P{}^2 = 3M\=I. \l{trPP}\f
Because we have
\e \=P{}^{(4)} = M^2\=I{}^{(4)}\not=0, \f
$\=P$ must be of rank 4, whence the inverse dyadic $\=P{}^{-1}$ exists. Operating \r{P2MI} by $\LL\=P{}^{-1T}$ we have 
\e 3\=P = M(\tr\=P{}^{-1})\=I - M\=P{}^{-1}. \f
Multiplying this by $\=P|$ and combining with \r{trPP} leaves us with the condition
\e (3\tr\=P -M\tr\=P{}^{-1})\=P = 8M\=I. \f
Since from $M\not=0$ it follows that the bracketed quantity cannot vanish, $\=P$ must actually be a multiple of the unit dyadic,
\e \=P = \A\=I,\ \ \ \ \A=\pm\sqrt{M}. \f
The same solution was already obtained by considering the skewon-less P medium with no principal part or the principal-less P medium with no skewon part.

\subsection{Pure-skewon media}

To find the general expression for the pure-skewon medium we may start from \r{Mprincc}, which defines the P medium with no principal part. The axion part is deleted by requiring vanishing of the trace, 
\e \tr\=M = 6P'{}^2 + 3P'(\#a|\%\B)=3P'(2P'+\#a|\%\B)=0.\f
If $P'=0$, according to \r{Mprincc}--\r{Mprincc1}, the medium bidyadic $\=M$ vanishes. Thus, $P'$ must be nonzero, and equal to $-(\#a|\%\B)/2$. We conclude that
\e \=P = \#a\%\B - \frac{\#a|\%\B}{2}\=I, \f
which yields the medium bidyadic
\ea \=M &=& (\#a\%\B- \frac{\#a|\%\B}{2}\=I)^{(2)T} \nonumber\\
&=& \frac{(\#a|\%\B)^2}{4}\=I{}^{(2)T}  - \frac{\#a|\%\B}{2}\%\B\#a\WW\=I{}^T \nonumber\\
&=& (\=B_o\WW\=I)^T,\fa
with
\e\=B_o = - \frac{\#a|\%\B}{2}(\#a\%\B - \frac{\#a|\%\B}{4}\=I),\ \ \ \ \tr\=B_o=0. \f

\subsection{Pure-principal media}

The most general P medium with only a principal component can be found by starting from the medium bidyadic of skewon-free media studied above of which \r{pureprinc} does not have the axion component. Thus, the medium bidyadic of any pure-principal P medium can be expressed as 
\e \=M = ( \%\B_1\#a_1+\%\B_2\#a_2)^{(2)}=  \%\B_1\%\B_2\WW\#a_1\#a_2,\ \ \ \ \#a_i|\%\B_j=0. \l{pureprinc1}\f
Actually, the same result can be derived by starting from the condition \r{05M1LLI}, 
\e \=M\LL\=I= \=P{}^{(2)T}\LL\=I = (\tr\=P)\=P{}^T - \=P{}^{2T} = -\=P{}^T|(\=P-\tr\=P\ \=I)^T=0. \l{Pprinc}\f
It is obvious that $\=P$ cannot be of full rank because multiplying \r{Pprinc} by $\=P{}^{-1T}|$ would yield $\=P = \tr\=P\ \=I$, whence $\tr\=P=4\tr\=P=0$ leads to $\=P=0$, a contradiction. Expressing \r{Pprinc} as
\e (\=P-\frac{1}{2}\tr\=P\ \=I)^2 =  (\=P_o-\frac{1}{2}P\=I)^2 = 4P^2\=I, \l{Pprinc1}\f
we can easily find that assuming $P\not=0$ will lead to $\=M=0$, whence $\=P_o$ must be a nilpotent dyadic which will eventually yield the result \r{pureprinc1}.

\subsection{Summary: Decomposition of P media}

In contrast to the Q media, none of the Hehl-Obukhov components of a P medium is absolutely necessary, since any one of the three components may be set to zero. However, the pure-principal P medium appears quite restricted since its medium bidyadic is required to have rank one. The six special cases of P media can be summarized as follows.

\begin{itemize}
\item $\=M_1\not=0,\ \=M_2=0,\ \=M_3\not=0$, skewon-free P media.
\e \=P=P(\=I -2(\#e_1\ve_1+ \#e_2\ve_2)),\ \ \ \=M= -P^2(\=I{}^{(2)T} -2(\ve_{12}\#e_{12}+ \ve_{34}\#e_{34})).  \f

\item $\=M_1\not=0,\ \=M_2\not=0,\ \=M_3=0$, axion-free P media.
\e \=P = \=P_o +P\=I,\ \ \ \ \=M = \=P{}_o^{(2)T} + P(\=P_o\WW\=I)^T + P^2\=I{}^{(2)T}, \ \ \ \ \tr\=P_o=0,\ \ P^2 = \frac{1}{12}\tr\=P{}_o^2.\f

\item $\=M_1=0,\ \=M_2\not=0,\ \=M_3\not=0$, skewon-axion P media. 
\e\=P=P\=I + \#a\%\B,\ \ \ \=M = P^2\=I{}^{(2)T} + P\%\B\#a\WW\=I{}^T,\ \ \ \ P+\frac{1}{2}\#a|\%\B\not=0. \f 

\item $\=M_1\not=0,\ \=M_2=0,\ \=M_3=0$, pure-principal P media.
\e \=P = \#a_1\%\B_1+ \#a_2\%\B_2,\ \ \ \ \=M = \%\B_1\%\B_2\WW\#a_1\#a_2,\ \ \ \ \#a_i|\%\B_j=0. \f

\item $\=M_1=0,\ \=M_2=0,\ \=M_3\not=0$, pure-axion P media.
\e \=P=P\=I,\ \ \ \ \=M = P^2\=I{}^{(2)T}. \f

\item $\=M_1=0,\ \=M_2\not=0,\ \=M_3=0$, pure-skewon P media.
\e \=P=P\=I + \#a\%\B,\ \ \ \=M = P^2\=I{}^{(2)T} + P\%\B\#a\WW\=I,\ \ \ \ P=-\frac{1}{2}\#a|\%\B. \f

\end{itemize}

\section{Additional remarks: Closure equations}

The densities and fluxes of energy and momentum are summarized in the Maxwell stress-energy dyadic 
\begin{equation}
\=T=\frac{1}{2}[\%\Phi\wedge(\=I\L\%\Psi)-\%\Psi\wedge(\=I\L\%\Phi)],
\end{equation}
mapping vectors to three-forms \cite{MDEM}. It can be shown that $\=T$ is invariant under the dual substitution
\e
\%\Psi \rightarrow \zeta \%\Phi, \qquad\%\Phi \rightarrow -\zeta^{-1} \%\Psi,\l{eq:reciprocity}
\f
also known as electric-magnetic reciprocity \cite{Hehl}. Evidently, the parameter $\zeta$ has to be nonzero. One can require that the medium bidyadic is also invariant under \r{eq:reciprocity}. A short calculation then establishes that dual substitution preserves $\=M$ if and only if:
\e
\=M|\=M=-\zeta^{2}\=I{}^{(2)T}, 
\f 
which is a special closure equation. Notably, there exist linear transformations involving $\%\Phi$ and $\%\Psi$ that are more general than \r{eq:reciprocity}, but still produce no change in the Maxwell stress-energy dyadic. In this context, the above reasoning eventually leads to the {\it closure equation}  
\e
\=M|\=M=\alpha\=I{}^{(2)T}\l{eq:closure}
\f
where the sign of the constant on the right-hand side is now open \cite{ThesisAlberto}. 

The invariants $\{\%\Phi\W\%\Phi, \%\Phi\W\%\Psi, \%\Psi\W\%\Psi\}$ describe the configuration of the electromagnetic fields. Imposing that the third invariant is proportional to the first, regardless of how $\%\Phi$ and $\%\Psi$ are chosen, yields another closure equation for $\=M$. Requiring that a relation of the form
\e
\%\Psi\W\%\Psi=\A\%\Phi\W\%\Phi,\l{eq:invariants}
\f
whose both sides equal a scalar multiple of $\ve_N$, is valid in the medium for any field two-form $\%\Phi$ for some scalar $\A$, 
we must have
\ea \#e_N|(\%\Psi\W\%\Psi-\A\%\Phi\W\%\Phi) &=& \%\Psi|(\#e_N\L\%\Psi)-\%\Phi|(\A\#e_N\L\%\Phi) \nonumber\\
&=& \%\Phi|(\=M{}^T(\#e_N\L\=M)-\A\#e_N\L\=I{}^{(2)T})|\%\Phi =0 \fa
valid for any two-form $\%\Phi$. Since the bidyadic in brackets is symmetric, it follows that the medium bidyadic $\=M$ must satisfy a condition of the {\it modified closure equation}
\e
\=M{}^{T}|(\#e_N\L\=M)= (\#e_N\L\=M)^T|\=M =\A\#e_N\L\=I{}^{(2)T}.\l{eq:closuremod}
\f

If the medium bidyadic is skewonless satisfying $(\#e_N\L\=M)^T = \#e_N\L\=M $, the ordinary and the modified closure equations obviously become the same. The solutions to this unique closure equation are the skewonless P and Q media derived in this work. As a matter of fact, the generic P and Q media are the solutions to \r{eq:closuremod} when the skewon part $\=M_2$ is not necessarily zero \cite{MDEM,ThesisAlberto}.

\section{Conclusion}

Since the most general electromagnetic medium requires 36 medium parameters for its definition, it appears necessary to study classes of media with lower number of parameters. In the present paper two previously defined classes of media were considered, each defined by 16 parameters in terms of four-dimensional bidyadic formalism. To further reduce the number of parameters, the two classes labeled as those of Q media and P media were decomposed in subclasses in a natural manner in terms of a decomposition introduced by Hehl and Obukhov. The number of special medium subclasses thus defined amounts six for both P and Q media with the exception that some of the subclasses were shown to be nonexistent. Since each of the subclasses was defined without relying on any basis, they share the property of the Hehl-Obukhov decomposition as being invariant in any affine transformation. It is outside the scope of the present paper to speculate on possible realizations of these different medium classes. 

\section*{Acknowledgements}

The authors would like to thank F.W. Hehl and W.-T. Ni for useful discussions. AF gratefully acknowledges financial support from the Gordon and Betty Moore Foundation.

\section*{Appendix: Two identities}

Starting from the bac-cab rule \cite{MDEM}
\e \%\Th\L(\#A\L\%\A) = -(\%\Th|\#A)\%\A + (\%\A\W\%\Th)\L\#A \l{PhiAA}\f
valid for any two-form $\%\Th$, one-form $\%\A$ and bivector $\#A$, let us assume that $\%\Th=\ve_N\L\#A$. When $\#A$ is not a simple bivector, we can define a vector basis so that $\#A=A_{12}\#e_{12}+ A_{34}\#e_{34}$. In this case we have $\%\Th=A_{12}\ve_{34}+ A_{34}\ve_{12}$ and $\%\Th|\#A= \ve_N|(\#A\W\#A)=2A_{12}A_{34}$. The last term of 
\r{PhiAA} can be expanded as
\ea (\%\A\W\%\Th)\L\#A &=& (A_{12}\%\A\W\ve_{34}+ A_{34}\%\A\W\ve_{12})\L(A_{12}\#e_{12}+ A_{34}\#e_{34}) \nonumber\\
&=& A_{12}A_{34}\%\A = \frac{1}{2}(\#A|\%\Th)\%\A. \fa 
Thus, \r{PhiAA} becomes
\e \%\Th\L(\#A\L\%\A) = -\frac{1}{2}(\%\Th|\#A)\%\A , \l{PhiLAA}\f
in the special case when $\%\Th=\ve_N\L\#A$. When the bivector $\#A$ is simple, we have $\%\Th|\#A=0$.

To find another identity involving the expression $\=S\WW(\#A\L\=I{}^T)$ we insert the expansions $\#A= \sum A_{mn}\#e_{mn}$ and $\=S=\sum S_{ii}\#e_i\#e_i$. Applying
\e\#e_i\#e_i\WW(\#e_{mn}\L\=I{}^T) = \#e_i\#e_i\WW(\#e_n\#e_m - \#e_m\#e_n) = \#e_{in}\#e_{im} - \#e_{im}\#e_{in} \f
and setting $m,n=1,2$ we obtain
$$ \ve_N\L\sum S_{ii}(\#e_{i2}\#e_{i1} - \#e_{i1}\#e_{i2}) = S_{33}(-\ve_{14}\#e_{31}-\ve_{24}\#e_{32}) +S_{44}(-\ve_{31}\#e_{41} +\ve_{23}\#e_{42}) $$
$$ = S_{33}(\ve_4\#e_3\WW(\ve_1\#e_1+\ve_2\#e_2)) -S_{44}(\ve_3\#e_4\WW(\ve_1\#e_1+\ve_2\#e_2))$$
$$= (S_{33}\ve_4\#e_3-S_{44}\ve_3\#e_4)\WW\=I{}^T $$ 
\e = (\ve_N\L(\#e_{12}\W(S_{33}\#e_3\#e_3+S_{44}\#e_4\#e_4)))\WW\=I{}^T. \f
Thus, we have
\e \ve_N\L\sum S_{ii}(\#e_{in}\#e_{im} - \#e_{im}\#e_{in}) = (\ve_N\L(\#e_{mn}\W\=S))\WW\=I{}^T, \f
whence we arrive at the identity
\e \=S\WW(\#A\L\=I{}^T) = \#e_N\L(\=B{}^T_o\WW\=I{}^T),\ \ \ \ \=B{}^T_o=\ve_N\L(\#A\W\=S). \l{BoI}\f

\end{document}